
%
%
\magnification=\magstep1
\hoffset=0.0 true cm
\voffset=0.0 true cm
\vsize=23.5 true cm
\hsize=17.0 true cm

\baselineskip=20pt
\parskip=0pt

\parindent=22pt
\raggedbottom

\def\pp{\noindent\parshape 2 0.0 truecm 17.0 truecm 0.5 truecm 16.5 truecm}

\def\pn{\par\noindent}

\def\lro{{ $log\rho_0$ }}
\def\ro{{ $\rho_0$ }}
\def\lrc{{ $log r_c$ }}
\def\rc{{ $r_c$ }}
\def\feh{ [Fe/H] }

\def\lr{{ $log R_{GC}$ }}

\def\z{{ $Z_{GP}$ }}
\def\r{{ $R_{GC}$ }}

\def\ie{{\it i.e.} }
\def\mv{ $|M_V|$ }

\def\etal{{\it et~al. }}
\def\lsim{\hbox{ \rlap{\raise 0.425ex\hbox{$<$}}\lower 0.65ex\hbox{$\sim$} }}
\def\gsim{\hbox{ \rlap{\raise 0.425ex\hbox{$>$}}\lower 0.65ex\hbox{$\sim$} }}
\def\out{{\it Outer }}

\def\inn{{\it Inner }}

\def\dis{{\it Disk }}
\def\ext{{\it Extreme }}

\def\gc{globular clusters }
\def\c{{$C$ }}

\null\vskip 1.0 truecm

\centerline{\bf DYNAMICAL FAMILIES IN THE }\pn
\centerline{\bf GALACTIC GLOBULAR CLUSTER SYSTEM}
\bigskip
\bigskip

\centerline {M. Bellazzini $^1$, E. Vesperini$^2$, F.R. Ferraro $^3$, and
F. Fusi Pecci $^3$}
\bigskip

\centerline{$^1$ {\it Dipartimento di Astronomia, Via Zamboni 33,
I-40126 Bologna, Italy. }}
\par\noindent
\centerline{$^2$ {\it Scuola Normale Superiore, Piazza dei Cavalieri 7, I-56126
Pisa, Italy.}}\pn
\centerline{$^3$ {\it Osservatorio Astronomico di Bologna, Via Zamboni 33,
I-40126 Bologna, Italy.}}\pn

\bigskip
\bigskip
\bigskip
\bigskip
\bigskip\noindent
\bigskip\noindent
\bigskip\noindent
To appear on: {\it Monthly Notices of the Royal Astronomy Society}
\bigskip\pn
\bigskip\pn
\centerline{BAP 10-1995-041-OAB}
\bigskip
\bigskip
\bigskip
\centerline{e-mail: bellazzini@astbo3.bo.astro.it}
\centerline{fax: 39-51-259407}

\vfill\eject

\centerline{\bf ABSTRACT}
\bigskip\noindent
The Halo Galactic Globular Clusters (GCs) lying inside the solar
circle (\inn, \r $< 8 Kpc$, \feh$<-0.8$) are shown to keep no record of the
strong correlation between GC absolute integrated magnitudes
and core parameters (concentration and central density)
clearly detected for the Halo GCs lying outside  this circle (\out)
as well as (at lower statistical significance) for those
belonging to the \dis population ($\feh >-0.8$).

We present here both observational evidences and theoretical
simulations supporting the hypothesis of a {\it primordial} origin for
such correlations and of an {\it environment-induced}
destruction of them if absent.

In particular, this hypothesis lead us to argue that \inn Halo GCs
have undergone a chaotic, environment-driven, dynamical evolution not
shared by the \dis GCs though the radial zone of the Galaxy occupied by
the two groups is nearly the same.

\bigskip\noindent
\centerline {\bf 1. INTRODUCTION}
\medskip\noindent
The dynamical evolution of \gc (GCs) is likely to be heavily affected by the
environment of the host galaxy.
In the Galactic GC system, environmental effects are due to
interactions between the GCs and the Galaxy, either via evaporation mostly
controlled by the Galactic tidal force at the cluster perigalacticon
(Aguilar 1993) or via the dynamical shocks the GCs undergo during
their passages through (or near) the main substructures of the Galaxy,
{\it i.e.} the disk, the bulge or big gas clouds.
All these mechanisms as well as dynamical friction can lead to
GC destruction (see Aguilar, 1993 for an extensive discussion).

However, the existing \gc have survived nearly for an Hubble time,
and the expected present-day destruction rate is $5\pm 3$
cluster per Gyr (Hut \& Djorgovski 1992). This implies that only
$\sim 3 \%$ of the whole Galactic GC system will disappear in the
next Gyr or so. As a consequence, the bulk of the existing Galactic globulars
consists of stellar systems which have continuously evolved,
for several core-relaxation times, toward higher and higher
concentrations and, in some cases, till  core collapse
(Djorgovski \& Meylan, 1994, hereafter DM94).

The time scales of such an evolution can be very different from cluster to
cluster and strictly depend on the efficiency of the various mechanisms
involved during the whole cluster lifetime.
The dynamical status of a globular cluster must therefore store information
not only about its ``intrinsic'' evolution but also about the effects
induced by the environment through which it has passed along its orbits.
A straightforward example of this kind of information is for instance supplied
by the evidence that all of the $\sim$30 clusters which are thought to have
undergone core collapse lie inside the inner 10 $Kpc$ from the Galactic
Center (Chernoff \& Djorgovski, 1989): this is indeed the proof that
environmental effects accelerate dynamical evolution of the globulars.

Djorgovski and Meylan (DM94) have presented the most relevant correlations
between position in the Galaxy and structural parameters for the whole
Galactic GC system, showing also clear hints on the existence of
the quoted ``memories'' of the past GC history.  In particular,
we recall that:

\item{1)} GC structural parameters (\rc --the core radius, \ro
--the central density, \c --the concentration)
are correlated between each other and with
the cluster integrated absolute magnitude $M_V$, in the sense that
central concentration increases with increasing cluster luminosity
(see also van den Bergh 1994).

\item{2)} GC structural parameters and positions in the Galaxy display
a good  correlation, in the sense that clusters far from the Galactic
Center tend to be looser.

\item{3)} $M_V$ and \r,  where \r is the GC distance from
the Galactic Center, are found to be nearly uncorrelated between each other
($r=0.11$), if one excludes from the sample the five most distant clusters
whose origin is uncertain (see discussion below).

\noindent
The basic aims of the present work are:

\noindent
\item{a)} to detect different dynamical evolutionary conditions
or trends in (properly selected) different groups of Galactic globulars;

\noindent
\item{b)} to evaluate the relative importance of primordial and evolutionary
mechanisms in each group.
\pn
The simple idea at the basis of this analysis is that the correlation
between concentration and integrated magnitude is mostly driven by internal
conditions, whilst the correlation between GC structural parameters
and, for example, \r or \z -- where \z is the height over the Galactic plane --
must be related to environmental effects on the  GC dynamical evolution.

Within this approach, \r and $M_V$ are considered to be {\it primary}
parameters, that is directly related to quantities settled at the very
origin of the globulars (\ie apogalactica and cluster masses, respectively).
Moreover, they are supposed to undergo little changing
and evolution throughout the life of a typical cluster (Murray \& Lin 1992).
Finally, since we are looking for evolutionary characteristics, our analysis is
focused on GC core parameters, which certainly are the most
sensitive to any dynamical evolution (Murray \& Lin 1992, DM94).

\bigskip\noindent
\centerline {\bf 2. DATA-SET AND DEFINITIONS}
\medskip\noindent
The data-sets adopted are those listed by Djorgovski (1993) and Trager
\etal (1993). Note that the absolute values of integrated magnitudes (\mv) are
used throughout the paper as they rank clusters nearly as masses do.
{}From the original entries (143), we have actually considered only 113
objects as for the other GCs important data were missing or they were
classified by Djorgovski (1993) to be of low quality.
Furthermore, because of their badly defined core
parameters,
we were forced to exclude from our sample
the GCs classified {\it "c"} and {\it "c?"} by Djorgovski (1993),
candidate members of the group of the so-called Post Core-Collapse Clusters
(PCC), (see, e.g., Lugger et al. 1995 for a recent
investigation on the core sizes of PCC clusters leading to values
significantly different from those provided by Trager \etal 1993).
The data concerning PCCs were used only in the computation of average
values for \feh, \mv and $|Z_{GP}|/R_{GC}$ (see Tab. 1) and in evaluating
correlations between parameters other than \ro, \rc or \c, as for example the
above quoted one between \r and \mv.

The final total sample has then been divided into four groups:

\noindent
\item{1.} {\it Disk Clusters}: having $\feh \ge -0.8$ (Zinn 1985).

\noindent
\item{2.} {\it Inner Halo Clusters}: having $\feh < -0.8$ and lying inside
 the solar circle, adopted to be at \r$=8 Kpc$.\pn

\noindent
\item{3.} {\it Outer Halo Clusters}: having $\feh < -0.8$ and lying
outside  the solar circle, but inside $40 Kpc$ from the Galactic Center.

\noindent
\item{4.} {\it Extreme Halo Clusters}: situated at distances larger
than $60 Kpc$ from the Galactic Center (as known, no cluster is found between
$\sim 40$ and $\sim 60 Kpc$ from the Galactic Center, Zinn, 1985; Armandroff,
1993).

\smallskip\noindent
The main characteristics of each group are presented in Table 1.
While the reasons to consider \dis and \ext clusters as singular families
are straightforward (Zinn 1985, 1993; Armandroff 1989), the partition of the
halo clusters in two subgroups at a fixed \r can seem rather arbitrary.
The aim was that of examining the dynamical status of two subgroups which
certainly spend most of their {\it life} in different environmental conditions
though having similar ({\it halo-like}) kinematical properties. Such a
claim can be supported for instance by the results of Chernoff \etal
(1986) who found that the impact of disk shocking on the GC system is
highly effective in the radial range $3 - 8 Kpc$ and also by those
presented by Weinberg (1994) who, on a quite different basis, suggests
that ``shock heating plays a defining r\^ole in the evolution of clusters
inside the solar circle''. Furthermore, setting the separation limit
at $8 Kpc$, we obtain a sample of halo  globulars (the \inn one) actually
sharing the same radial zone of the Galaxy as the \dis ones
(with the only exception of the disk cluster Pal 8 at $R_{GC}=20.9 Kpc$),
allowing thus
a direct ``local'' comparison. Finally, it has to be stressed that we
have verified that the results we present here are essentially unchanged
if the limit is moved $1-2 Kpc$ inward or outward.

\medskip\noindent
\centerline{\bf 3. OBSERVATIONAL EVIDENCES}
\medskip\noindent
In order to decide whether a correlation between two physical quantities
is present or absent in a specified sample we use the statistically robust
non-parametric Spearman rank correlation coefficient $s$ (see Press \etal
1992).
In calculating this coefficient the absolute
values of two given parameters (whose distribution laws are often unknown)
are substituted by the {\it ranks} of their values. The main advantage is
that the distribution function of the ranks ({\it i.e.} 1,2,3,4...) is
perfectely known (uniform distribution) and one can safely test
the significance of any deviation of the measured $s$ from an expected value
(say $0$, {\it i.e.} no correlation).
\pn
In particular, given two vectors {\bf a} and {\bf b} having dimension $N$,
the parameter
$$t=s{\sqrt{{N-2}\over{1-s^2}}}$$
is distributed approximately as a Student's distribution with $N-2$ degrees
of freedom, indipendently of the original distribution of {\bf a} and {\bf b},
and tests the null hypothesis ($H_0$) that the two vectors are uncorrelated
(Press \etal 1992).

An analogous test of significance can be performed also for the ordinary
linear correlation coefficient --the Pearson's $r$ (Taylor 1984,
Press \etal 1992)--
but it is valid only if the distributions of {\bf a} and {\bf b} jointly
form a {\it binormal Gaussian} distribution around their mean values
(Press \etal 1992).

For clarity and completeness, we report in any plot the values of
$r$ (which can at lest be regarded as an indication), $s$ (which
measures the {\it strength} of a correlation), and $x_s$ (\ie the percent
probability that the two considered vectors are uncorrelated).
In particular, the quantity $1-x_s$ can be regarded as the
quantitative confidence level of a given correlation.
Note that the use of ranks instead of absolute values prevents
from the any possible {\it overweighting} of outliers data which
can significantly affect the ordinary linear correlation coefficient $r$.
\pn
In conclusion, the presence/absence of any correlations in the data (and their
strength) is evaluated in a safe non-parametric way. In particular,
we will consider a correlation {\it significant} when $x_s\le 5\%$
and {\it very significant} when $x_s\le 1\%$.
\pn
The correlations detected for the GC groups (\dis, \inn and \out )
in the planes formed by the considered structural parameters
versus \lr and \mv are presented in Table 2. For each sample
the measured $s$ and $x_s$ values are reported for each couple of
considered parameters.

\smallskip\noindent
{\it 3.1. Trends with \r}
\smallskip\noindent
The main indication emerging from inspecting the first three columns of
Table 2 is that
the dynamical status of the \out clusters, in spite of the large radial
range spanned by the sample ($8--40 Kpc$), is practically unaffected by their
location within the Galaxy, while both the \dis and the \inn clusters show
some correlation. To give a view of the different behaviours of the various
sample, we report in Fig. 1 the distribution of the data points in the plane
\lr -- \lro for the \out, \inn and \dis samples, respectively. The full lines
represent the linear regression fits to the data.
\pn
The difference between the \out group and the others is also testified to
by the fact that the \out sample as a whole displays a significantly different
distribution of concentrations than the \inn and \dis ones, centered on
lower \ro and \c values. Moreover, this global evidence is consistent with the
result of Chernoff \& Djorgovski (1989) who showed that the fraction of PCC
clusters increases with decreasing \r, and it gives further support
to the idea that the evolution of GCs is strongly affected by the
Galactic environment, being faster for clusters close to the Galactic Center
where the tidal field of the Galaxy is much stronger.

\smallskip\noindent
{\it 3.2. Trends with \mv}
\smallskip\noindent
Turning to Table 2, columns 4--6, it is evident that
the \out clusters  show a clear-cut and {\it very significant} correlation
between \mv and structural parameters in any of the examined planes.
Also the \dis clusters display good correlations, though less evident and
significant. On the contrary, in spite of the strict similarity in the
radial range actually spanned with the \dis sample, the \inn clusters
do not show any statistically significant correlation.  Hence, we
can point out here the existence of a striking difference between the
two subgroups of {\it halo} Galactic GCs.
\pn
In this respect, it is particularly interesting to consider the plane \mv
--\lrc,
where both quantities are {\it direct observables}, and where
the \inn clusters display (if anything) a very weak correlation just in the
{\it opposite} sense than that of the other two groups (\ie there is no
possible doubt on the {\it absence} of this correlation in the \inn
sample).
It is also
worth recalling that while  the correlation is clearly present in the
\out sample and possibly in the \dis one, the analysis of the
global sample (DM94, van den Bergh 1994) shows only a {\it weak} correlation
in this specific plane, clearly masking the differences associated to the
various sub-groups.

\medskip
In order to present the results in more detail and to demonstrate their
statistical significance, we show in Figure 2 the actual distributions of
the clusters included in the three samples in the \mv-\lro plane.
The scale is the same in any panel and
covers the whole range spanned by the considered Galactic GCs.
The full lines are linear regression fits to the data;
$s$, $x_s$ and $r$ are the statistical parameters described above.
\pn
By inspecting the two {\it upper panels} (concerning the \out and \inn
clusters), one could argue that the absence of any clear-cut correlation
in the \inn sample is simply due to the narrower range spanned in both
\mv and \lro than the \out sample. However, as shown in the right-bottom panel,
this is not the case. In fact, by extracting from the \out group a subsample
(\out-B) sharing exactly the same parameter ranges as spanned by the
\inn globulars  one gets a still highly significant correlation
($>99 \%$ confidence level).
Finally, it is worth noting that the same kind of analysis
applied to the \mv- \c ~~and \mv -\rc planes would yield fully consistent
results.

In conclusion, independently of any apparent bias, it seems
quite evident that within the \out sample it is possible to detect a clear-cut
trend which is not present in the \inn one. As discussed below, this
fact may have important implications in the understanding of
GC formation and dynamical evolution.

\smallskip\noindent
{\it 3.3. A bivariate analysis}
\smallskip\noindent
Before concluding this Section it may be worth reporting the results
obtained by applying a bivariate analysis to the three groups
to find possible linear combinations of the type:
 $$Q=a(|M_V|+blog R_{GC})$$
which could optimize significantly the correlation coefficient with respect
to the best monovariate analysis, and where $Q$ can be in turn \lrc, \lro, \c,
respectively. The results so obtained are different for the three groups.

In the \out sample, no significant difference has been detected with respect
to the monovariate framework, being each considered plane actually
dominated by the strong ranking induced by \mv.

The \inn group is slightly affected, in the sense that the contemporary
consideration of \r and \mv marginally improves the correlations.

The \dis sample shows a clear-cut improvement in the correlations,
evidence for an intrinsic bi-parametric dependence. As can be seen
in Figure 3 where  the \dis clusters are plotted in the plane \lro
versus $|M_V|-3.5log R_{GC}$, $s$ increases by
$\sim0.2$ with respect to the monovariate case (the plane \mv-\lro),
and the significance level is very high. In particular, the relationship
is quite remarkably narrowed, and this is a further confirmation
that \dis clusters recorded the ``signals'' impressed by both \mv and \r.

\smallskip\noindent
{\it 3.4. The Extreme clusters}
\smallskip\noindent
The poorness of the \ext sample (6 clusters) prevents any significant
statistical analysis of the kind performed for the other considered groups.
Because of their position and chemical properties many authors have argued
for an independent origin of these clusters with respect to the Galactic GC
system (Harris 1976, van den Bergh 1983, Zinn 1985).
Fusi Pecci \etal (1995) have shown that four of them
(Pal 14, Pal 3, Pal 4 and Eridanus), which span a narrow
metallicity range, have plunging orbits and lie on a plane in the sky
passing through the Galactic Center (see also Majewski 1994).
\pn
Moreover, while no significant difference is found between the luminosity
functions of the \dis, \out and \inn samples, a KS-test shows that the
probability that the \ext  and the $\inn+\out$ samples are drawn from the
same parent population in \mv is less than $0.3 \%$. Note that
if the same KS-test is performed comparing just the \ext and \out samples
the probability increases only up to  $\simeq 0.6\%$, while
for example for the \inn and \out samples the probability is $\sim81\%$.
Hence it seems reasonable to isolate this small group of distant
globulars and consider them separately.

For what concerns the specific analysis carried out above,
5 of these \ext clusters lie on the same relations as found for the
\out globulars in the planes "luminosity {\it vs.} core concentration", while
only one, NGC 2419, greatly deviates, being NGC 2419 very luminous and sparse.
This fact is not totally at odds with the general trend as NGC 2419,
very bright and loose, is anyway much denser than the other, less massive,
\ext clusters.

\medskip\noindent
\centerline{\bf 4. THEORETICAL SIMULATIONS}
\medskip\noindent
The observational evidences presented in Sect. 3 ask for interpretation
and modeling. In this section we describe the main results of a
theoretical investigation on the evolution of GC systems
(see also Vesperini 1994, 1995a,b for further details) which may add
information useful to properly interpret the presence or lack of
correlation between the various quantities in the considered GC samples.
In particular, the main goal of the present simulations is that of
determining the r\^ole of primordial conditions and evolutionary processes
in establishing the observed correlations. Due to the uncertainties on the
differences between halo and disk clusters, particularly as far as age and
interaction with the disk are concerned,
we have limited our analysis to a population of halo clusters.

\smallskip\noindent
{\it 4.1 The model assumptions and parameters}
\smallskip\noindent
In order to follow the evolution of GCs with a large set of
different initial values of mass and concentration, and located at
different distances from the Galactic Center, we have used a method
introduced by King (1966) and later on applied by Prata (1971a,b), Chernoff
\etal (1986), Chernoff \& Shapiro (1987), based on the assumption that GC
evolution can be described, at least until the onset of gravothermal
instability, as a sequence of King models with time evolving concentration,
mass
and radius.  Here, we give just a brief description of the method (for further
details see the above references).

Given  the total energy $E_{tot}$, the total mass $M$, and the truncation
radius $r_t$,  a King-model describing an individual GC is univocally
determined.
We shall assume that the truncation radius is determined by the tidal field
of the Galaxy (see, e.g., Spitzer 1987) and defined as
$$ r_t=\left(M\over 3M_g\right)^{1/3}R_{GC},\eqno(1)$$
where $R_{GC}$ is the radius of the orbit of the GC around the host
galaxy and $M_g$ is the mass of the galaxy contained inside that radius.
For simplicity, the cluster orbits are taken to be circular and lying
on planes perpendicular to the plane of the galactic disk.

After a given time interval $\Delta t$, the change of mass and
energy  produced by evolutionary processes can be calculated and a new tidal
radius can then be determined by the tidal limiting condition (eq. 1). The
three
new values of mass, energy and radius identify a new King-model with different
concentration and different scale parameters describing the cluster at time
$t+\Delta t$. The total change in mass and energy is given by the sum of the
variations of these quantities due to all the processes taken into
consideration in the study.

In our investigation we have included the effects on the mass and the energy of
the cluster due to  disk shocking and internal relaxation  (see Chernoff \etal
1986 and Vesperini 1994, 1995a,b  for further details).
As for the model of the disk of the Milky Way, we have adopted the same one
used in Chernoff \etal (1986) obtained by fitting  with a two-component
isothermal  model the acceleration along the $z$ direction in the solar
neighbourhood determined by Bahcall (1984)
$$
K_z(R_0,z)=\sum_iK_i\hbox{tanh}\left({z\over z_i}\right) \eqno(2)
$$
with $R_0=8~ Kpc$, $K_0=3.47 \times 10^{-9} \hbox{cm s}^{-2}$, $z_0=175~pc$,
$K_1=3.96 \times 10^{-9} \hbox{cm s}^{-2}$, $z_1=550~pc$.
Acceleration varies with the Galactocentric distance according to the surface
density that, according to the Bahcall-Schmidt-Soneira (1982) model for the
Galaxy we will adopt in our investigation, falls off exponentially with an
exponential scalelength $h=3.5 Kpc$.
Clusters are assumed to rotate with a constant velocity $V=210 \hbox{km
s}^{-1}$ corresponding to a constant ratio $M_G/R_{GC}=10^7
\hbox{M}_{\odot}/\hbox{pc}$.

\noindent
The evolution of each cluster is followed until one of
these three conditions is satisfied:

\item{1.}  $W_0 > 7.4$ --limit for the onset of gravothermal catastrophe
($W_0$ is the dimensionless central potential). Both analytical studies and
numerical integrations of the Fokker-Planck equation  (Katz
1980, Wiyanto \etal 1985) indicate that King models with $W_0$
larger than this value are unstable against gravothermal catastrophe.
Gravothermal collapse is driven by a thermodynamic instability and can not be
described by the above method.

\item{2.} $W_0<0.05$ --under this value the system is conventionally considered
to be dissolved (Chernoff \& Shapiro 1987).

\item{3.} $t=1.5 \times 10^{10}$ years (Hubble time). The present era has been
reached.

\noindent
Our final sample of globular clusters include only those clusters that can
still be represented as King models after one Hubble time.

The main limitations of our model are the impossibility of following the
evolution of clusters after the onset of gravothermal catastrophe and the fact
that we consider only clusters on circular orbits.
In a future
work we will extend our analysis by including clusters on eccentric orbits but
it
must be noted that this leads to a significant increase of the parameter space
of
initial conditions.
Restriction to circular orbits implies that some caution is necessary in
comparing theoretical results with observational data particularly for what
concerns trends with galactocentric  distance, as no mixing of orbits is
present in our model. On the other hand we point out that a detailed comparison
of the theoretical results from simulations with observational data is beyond
the scope of our analysis; the main goal is rather that of establishing whether
evolutionary processes can modify given initial conditions for a GC
system toward the observed properties or it is necessary to apply to
initial conditions to explain them.
\smallskip\noindent
{\it 4.2 Evolution of a globular cluster system}
\smallskip\noindent
In order to start our simulation to follow the evolution of a GC system
we need to set the initial conditions: the distribution of orbital
parameters, the initial mass function of the GC system, the distribution of
initial concentrations of the GCs, and the spatial distribution of
the GCs in the Galaxy.

The present  knowledge of the real initial conditions of the Galactic GC
system is rather poor, and it is not clear to what extent the present
observational properties of the GCs keep memory of the initial ones.
Thus we have investigated different initial conditions and here we describe the
results of two of the simulations we have carried out, pointing out the main
points of interest  connected with the issues addressed in the paper (see
Vesperini 1994, 1995b for the results concerning other properties of globular
cluster systems).

The results described below have been obtained starting from a sample of
1000 clusters located on circular orbits with Galactocentric distances
between 1 and 20 $Kpc$, and with a spatial distribution in the Galaxy such
that the number of GCs per cubic kiloparsec is $N(R_{GC}) \propto
R_{GC}^{-3.5}$ according to what observed for $R_{GC}>4 Kpc$ (see, e.g., DM94).
The initial mass function has been chosen
to be a truncated  power-law, $f(M)\propto M^{-2}$, with lower cut-off
$M_{low}=10^{4.5} M_{\odot}$ and upper cut-off $M_{up}=10^{6} M_{\odot}$.
As for the distribution of concentrations, $C$, we have investigated two
possibilities: the first ({\it ~run I~}) with $C$ and $\log M$ {\it initially
correlated}
$$
C=-2.0+0.6\log M \eqno(3)
$$
chosen on the basis of the properties of the more massive and distant clusters
(but it is important to point out that the qualitative conclusions of the
theoretical analysis do not strongly depend on the exact form of the initial
relation between $C$ and $\log M$). The second ({\it ~run II~}) starting with
$C$ and $\log M$ {\it uncorrelated} in order to test the hypothesis that the
correlation between concentration and mass is due to evolutionary processes
(Djorgovski 1991; DM94). In {\it run II}, the initial distribution of $C$
is uniform in the range $[0.2-1.5]$.

Figure 4a,b show the final $C-\log M$ plane for the two simulations actually
carried out.

\item{\it run I}: starting with $C-\log M$ correlated. The correlation is
essentially preserved even though with a large spread (see Fig. 4a) due to the
effects of evolution causing the concentration to decrease or increase
depending
on the initial condition of the system; evolutionary processes modify the slope
of the relation and give rise to the observed scatter.

\item{\it run II}:  starting with $C-\log M$ uncorrelated. The final sample
does not show any trend between the two quantities  (see Fig. 4b),
confirming the skepticism about the possible r\^ole of evolutionary processes
in giving rise to the observed correlation.

\noindent
Figures 5a,b show  the $C-\log M$ plane for the final sample resulting from the
{\it run I} for inner ($R_{GC}<8 Kpc$) and outer ($R_{GC}>8 Kpc$) clusters: it
is evident that the initial correlation between concentration and mass  of the
clusters is more efficiently destroyed in the {\it Inner } regions where
evolutionary processes are more efficient, while for clusters  in the {\it
Outer } regions the correlation is stronger since the memory of the initial
conditions is preserved. The Pearson's, $r$, and the Spearman's, $s$,
correlation coefficients for the final  $C-\log M$ relationship for the two
samples have the same trend present in observational data being $r=0.95$,
$s=0.94$ for the \out clusters and $r=0.57$, $s=0.56$ for the \inn clusters.

Also for the $C-\log R_{GC}$ relationship the final result of the simulation
({\it run I}) shows the same trend present in the observational data. Initial
conditions are such that there is no correlation between concentration and
galactocentric distance neither for {\it inner} nor for {\it outer } clusters.
In the final sample,  inner clusters display
a trend between $C$ and $\log R_{GC}$ in the sense of clusters having higher
concentrations at smaller galactcocentric distances, with  correlation
coefficients ($r=-0.60$, $s=-0.64$).
No correlation between $C$ and $\log R_{GC}$ is present ($r=0.17$, $s=0.24$)
for outer clusters also at the end of the simulation.
The trend for {\it inner } clusters to have higher concentrations as
the galactocentric distance decreases is completely originated by evolutionary
processes partly as a consequence of the environmental effects accelerating
the
evolution of clusters located in the inner regions of the Galaxy and partly as
the result of a selective depletion of low-concentration clusters.
\pn
Trends similar to the observed ones are obtained also for the other
relationships concerning core parameters-mass and core
parameters-galactocentric distance.

Before closing, it is important to point out that the good qualitative
agreement between theoretical and observational results
has been obtained in spite of our
conservative assumption on the range of Galactocentric distances for \out
clusters in the theoretical sample $[8-20] \hbox{Kpc}$ compared with the
observational one $[8-40]\hbox{Kpc}$. By including larger values of $R_{GC}$,
the fraction of clusters in the sample of the \out GCs preserving memory of
initial conditions would increase, making the differences between \inn
and \out globulars even larger.

\medskip\noindent
\centerline{\bf 5. DISCUSSION}
\medskip\noindent
{\it 5.1. A ``primordial'' origin for the correlation \mv--core parameters}
\medskip\noindent
The new significant result presented here is that the relation standing
between absolute luminosity (\ie total mass) and core parameters
(\ie concentration and central density) of Galactic GCs appears to
be mostly evident in the group of clusters lying in an external region
of the halo (\out clusters), whilst it is substantially absent in the inner
halo objects (\inn clusters).

The very high statistical significance of these observed relations
suggests that, in spite of the fact that the members of \out group seem to
have preferentially plunging orbits (van den Bergh 1994), they were
poorly disturbed by evolutionary dynamical processes.
Moreover, these clusters do not show any ranking in concentration
induced by their distance from the Galactic Center, and this weakens the
possibility that the detected trend could have an environmental origin.
\pn
Taking into account the above evidences and the results of our
theoretical simulations, concerning the origin of the detected correlation
{\it luminosity-concentration} we are led to conclude that:

\item{} The correlation has a primeval origin, in the sense that more massive
clusters were {\it born} more centrally concentrated. In our view,
evolutionary processes are likely to have the effect of destroying the
correlation established at the time of cluster formation. In fact,
both the observational data and the results of our theoretical investigation
show that the studied correlations are much weaker for clusters in the
inner regions of the Galaxy, where all evolutionary processes are more
efficient. In this respect, it is interesting to note that a similar
correlation
between concentration and luminosity holds for dwarf elliptical galaxies (see,
e.g., Fig. 10b in Binggeli \etal 1984, Fig. 11 in Ichikawa \etal 1986) for
which evolutionary processes are likely to be irrelevant.
The  confirmation  of this working hypothesis would obviously be of great
relevance in constraining the physical processes playing a r\^ole
in the formation of these stellar systems.
\pn
Alternatively, one can hypothize that there is an {\it internal} mechanism
accelerating the evolution
of the more  massive clusters and making them  more and more
centrally concentrated.
Such a mechanism has necessarily to be {\it internal} because the
mass of the cluster is (at first order) determined only by the {\it
primordial} conditions at the time of its formation.
As a consequence, this kind of evolution has to affect also an ideal isolated
cluster, \ie in the quoted range of parameters a larger mass in some way
accelerates two-body relaxation and, in turn, the evolution of the
cluster toward higher concentrations.
Still, for a given size of the system,  two-body relaxation time
is proportional to $M^{1/2}$, and this runs just in the opposite direction
needed to explain the observed trend via evolutionary effects.

\medskip\noindent
{\it 5.2 Comments against the ``environmental'' origin}
\medskip\noindent
The sample Galactic globulars here considered is necessarily made just of
{\it survived} GCs. One might thus imagine that any relation involving
cluster structural properties could well be originated by selective depletion
processes within the original GC system.
\pn
However, while it is acceptable that the loosest massive clusters
were somehow disrupted, it seems very difficult to understand which process
could have destroyed the more massive tail of the "loose cluster" distribution,
leaving the lighter population still alive.

In order to explain the global \c-\mv and \lro-\mv trends, DM94  have
invoked "... a differential survival effect, with more massive clusters
surviving longer and reaching more evolved dynamical states".
We are however skeptical about this scenario mainly because, since present-day
clusters {\it have all survived till now}, the only way to actually allow
massive clusters to evolve longer is to postulate that less massive GCs are
significantly younger than the others, which is so far not confirmed by
any observational evidence. Moreover, it is unlikely that the high-mass,
low-$C$ GCs in the outer regions of the Galaxy could evolve enough to
depopulate completely the high-mass, low-$C$ part of the $log M - C$ plane.
As shown also in the results of our theoretical simulations  ({\it ~run II~}),
if present in the primordial GC system, high-mass, low-$C$ clusters should
still exist at present.

Finally, assuming that the effect of disk shocks could be important
also at large Galactocentric distances, one could imagine that the
GCs more massive than a certain critical value ($M_*$) are induced by
the  shock events to contract their cores, while those less massive than $M_*$
are induced to expand. This is exactly the way in which disk shocks are
believed
to act, but the critical parameter is found to be the {\it concentration} and
not the {\it mass} (Chernoff \etal 1986; Weinberg 1994). So this is a viable
explanation for the detected trend only if a {\it primordial}
"mass-concentration" coupling is postulated.

\medskip\noindent
{\it 5.3 Differential impact of disk shocks in the inner regions of the Galaxy}
\medskip\noindent
Another observational evidence
seems to yield a further important indication as the impact of environmental
effects are apparently different on different groups of clusters,
even if located within the same region of the Galaxy.
\pn
In fact, the \dis clusters, which actually occupy the same radial zone
of the Galaxy as the \inn globulars, keep record of the internal
processes (being ranked by the quoted linear combination of \mv and \lr),
while the \inn clusters do not.
What is the cause of such a different behaviour for the two groups?

In our view, the most viable explanation is that, due to their very different
orbits, the \inn globulars experience stronger interactions with respect to
\dis clusters, during the passage
through the same Galactic environment.
\dis clusters are orbiting circularly near the plane of the Galaxy and
they are rotating with approximately the same velocity of their stellar
neighbours (Zinn 1985, Armandroff 1989, Majewski 1993), so minimizing the
shocking effect of the Galactic Disk (Aguilar, 1993).
On the contrary, the \inn clusters are known to have a small net rotation
compared to that of the Galactic Disk (Armandroff 1989, Zinn 1993),
and they have a distribution of the heights on the Galactic Plane
spread out over a very wide range and suggestive of orbits
incident on the Galactic Plane with large inclinations.
Due also to the rapid Galactic crossing time (Chernoff \& Weinberg 1990),
they had probably undergone a great number of passages through the densest
parts of the Galactic disk, suffering several episodes of accelerated
evolution (DM94) which erased any memory of the ranking settled by
the {\it primordial} formation processes.

To get a direct confirmation of the different orbital characteristics
actually displayed by the two groups, we have performed
a straightforward test of flattening comparing the distributions of
the parameter $(|Z_{GP}|/R_{GC}$) in the two (\dis and \inn)  groups.
\pn
A KS-test rejects the hypotesis that the two samples
are drawn from the same population at $99.9\%$,
the \dis sample being significantly flatter on the Galactic
Plane than the \inn one.  Moreover, looking at the data presented in
Table 1, it can also be noted that the \dis sample is the only one
showing a $<|Z_{GP}|/R_{GC}>$ value hardly compatible with a uniform
spherical distribution around the Galactic Center.

Clues on a possible major effect of disk shocking on Galactic globulars
have been recently pointed in a previous paper (Bellazzini \etal 1995),
where we showed (based on the observed frequency of Low Mass X-ray Binaries
in the Galactic GCs) that the increasing of the slope of GC IMF with
increasing \r and \z (Djorgovski, Piotto and Capaccioli, 1993),
if real, cannot be primordial but it can be settled by the selective depletion
of the lighter stars from the halo of the clusters during their passage
through the densest internal parts of the Galactic Disk, as already
envisaged by the same latter authors.

\medskip\noindent
\centerline{\bf 6. CONCLUSIONS}
\medskip\noindent
The results presented here add new observational evidences on the impact of
disk shocks on Galactic GCs. Further, among the various correlations
detected in different GC groups between GC integrated absolute
magnitudes and intrinsic structural parameters, it is now possible
to discriminate those likely to be primordial and those likely to
result from evolution.  A proper set of theoretical simulations
on the evolution of a GC system reproduce the observed trends and
support our claims.

\noindent
The main conclusions drawn from the present analysis are:

\medskip\noindent
\item{1)} Considering the whole set of Galactic GCs, the already known
evidence that GC concentration increases with increasing absolute luminosity
is confirmed. Our analysis shows in addition that the correlation is
very strong for clusters located in the outer regions of the Galaxy,
where the tidal field is weaker, and almost absent (probably because
destroyed by evolution) in the inner halo. This lends strong support
to the hypothesis of a primeval origin of the observed correlation,
where it exists.
In this respect, the results of theoretical simulations on the evolution
of GC systems we have carried out show that evolution tends to erase the
primordial concentration -total luminosity correlation  and not to create
it as claimed by some authors (Djorgovski 1991, DM94).

\medskip\noindent
\item{2)} The \inn clusters show a much better correlation between core
parameters and Galactocentric distance, \r, than the \out globulars.
This is likely the result of evolutionary processes, more efficient as
\r decreases. The \out clusters are probably closer to their primeval status,
and do not show any significant trend with \r.

\medskip\noindent
\item{3)} The impact of the environment has been apparently very different
for two groups of clusters (\inn and \dis) actually sharing the same radial
zone of the Galaxy. In fact, \dis clusters display a well defined
correlation, whilst \inn globulars do not.
A plausible mechanism able to cause such a difference is disk shocking which
had a major impact on the inclined, high \z orbit of the \inn
clusters and a neglegible one on the very flat, rapidly
rotating \dis system.

\medskip\noindent
\item{4)} The galactic environment plays apparently a major r\^ole in the
dynamical evolution of GCs only in the inner $8-10 Kpc$ from the Galactic
Center, while the population presently found outside this circle seems
to be almost unaffected by environment-induced effects, preserving very
well their primeval intrinsic properties.

\bigskip\noindent
\bigskip\noindent
\vfill\eject

\centerline{\bf References}
\bigskip

\pp Aguilar, L. A., 1993, in Galaxy Formation: The Milky Way Perspective,
ASP Conf. Series, 49, 155

\pp Armandroff, T. E., 1989, AJ, 97, 375

\pp Armandroff, T. E., 1993, in Galaxy For\-ma\-ti\-on: The Mil\-ky Way
Per\-sp\-ect\-i\-ve, ASP Conf. Series, 49, 167

\pp Bahcall, J.N., 1984, ApJ, 287, 926

\pp Bahcall, J.N., Schmidt, M., \& Soneira R.M., 1982, ApJ, 258, L23

\pp Bellazzini, M., Pasquali, A., Federici, L., Ferraro, F.R., \&
Fusi Pecci, F., 1995, ApJ, 439, 687

\pp Binggeli, B., Sandage A., \& Tarenghi M., 1984, AJ, 89, 64

\pp Chernoff, D., \& Djorgovski, S.G., 1989, ApJ, 339, 904

\pp Chernoff, D., \& Shapiro, S., 1987, ApJ, 322, 113

\pp Chernoff, D., \& Weinberg, M., 1990, ApJ, 351, 121

\pp Chernoff, D., Kochanek, C., \& Shapiro, S., 1986, ApJ, 309, 183

\pp Djorgovski, S.G., 1991, in Formation and Evolution of Star Clusters,
ASP Conf. Series, 13, 112

\pp Djorgovski, S.G., 1993, in Structure and Dynamics of Globular Clusters,
ASP Conf. Series, 50, 373

\pp Djorgovski, S.G., \& Meylan G., 1994, AJ, 108, 1292 (DM94)

\pp Djorgovski, S.G., Piotto, G., \& Capaccioli, M., 1993, AJ, 105, 2148

\pp Fusi Pecci F., Bellazzini, M., Ferraro, F.R., \& Cacciari, C., 1995, AJ,
in press

\pp Harris, W.E., 1976, AJ, 81, 1095

\pp Hut, P., \& Djorgovski, S.G., 1992, Nature, 359, 806

\pp Ichikawa, S., Wakamatsu, K., \& Okamura, S., 1986, ApJS, 60, 475

\pp Katz, J., 1980, MNRAS, 190, 497

\pp King, I. R., 1966, AJ, 71, 64

\pp Lugger, P.M., Cohn, H.N., \& Grindlay, J.E., 1995, ApJ, 439, 191

\pp Majewski, S. R., 1993, Ann. Rev. Astron. Astrophys., 31, 575

\pp Majewski, S.R., 1994, ApJ, 431, L17

\pp Murray, S.D., \& Lin, D.N.C., 1992, ApJ, 400, 265

\pp Prata S.W., 1971a, AJ, 76, 1017

\pp Prata S.W., 1971b, AJ, 76, 1029

\pp Press, W. H., Teukolsky, S. A., Vetterling, W. T., \&
Flannery, B. P., 1992,
Numerical Recipes, 2nd ed., Cambridge University Press, New York, USA

\pp Spitzer L., 1987, 'Dynamical Evolution of Globular Clusters', Princeton
University Press

\pp Taylor, J.R., 1982, An Introduction to Error Analysis, U.S.B, Oxford
University Press, Mill Valley, CA, USA

\pp Trager, S.C., Djorgovski, S.G., \& King, I.R., 1993, in Structure and
Dynamics of Globular Clusters, ASP Conf. Series 50, 347

\pp van den Bergh, S., 1983, ApJS, 51, 29

\pp van den Bergh, S., 1994, ApJ, 435, 203

\pp Vesperini, E., 1994, Ph.D. thesis, Scuola Normale Superiore, Pisa, Italy

\pp Vesperini, E., 1995a,b, in preparation

\pp Weinberg, M., 1994, AJ, 108, 1414

\pp Wiyanto,  P., Kato S., \& Inagaki, S., 1985, PASJ, 37, 715

\pp Zinn, R.J., 1985, ApJ, 293, 424

\pp Zinn, R.J., 1993, in The Globular Cluster-Galaxy Connection, ASP
Conf. Series, 48, 38

\parskip=5pt

\vfill\eject
\centerline{\bf Figure Captions:}

\bigskip
\noindent{\bf Figure 1.} Distributions of the three cluster
samples ({\it Outer, Inner} and {\it Disk}) in the plane \lr- \lro.
The statistical parmeters $s$, $x_s$ and $r$ are described in Sect. 3.
The full lines represent the best fit linear regression loci.

\bigskip
\noindent{\bf Figure 2.} Distributions of the three cluster
samples ({\it Outer, Inner} and {\it Disk}) in the plane \mv- \lro.
The statistical parameters $s$, $x_s$ and $r$ are described in Sect. 3.
The full lines represent the best fit linear regression loci.
The low-left panel shows the subsample of \out clusters spanning
the same range of the observable plane as the {\it Inner} globulars.
As can be seen, the correlation present in this subsample is
still significant ($>99.9 \%$) (see Sect. 3.2).

\bigskip
\noindent{\bf Figure 3.} The clusters membering the {\it Disk} group are
plotted in the plane $|M_V|-3.5log R_{GC}$ -- \lro. The correlation coefficient
is enhanced by $\sim 0.2$ with respect to the monovariate case \mv-- \lro.
As can be seen, the relation is impressively narrowed with respect to
the monovariate case (Figure 2). This is a further indication
that {\it Disk} clusters keep memory of the ``signal'' impressed by
both \mv and \r.

\bigskip
\noindent{\bf Figure 4.} The final cluster distributions in the \c-$log M$
plane for the simulations carried out in {\it run I} (panel {\it a}) and
{\it run II} (panel {\it b}).
In both cases the imposed correlation ({\it run I}) and non-correlation
({\it run II}) were preserved till the present epoch (see Sect. 4).
The line drawn in panel {\it a} indicates the relationship
between $c$ and $\log M$ assumed for the initial conditions.

\bigskip
\noindent{\bf Figure 5.}
The final cluster distributions in the $C-\log M$ plane from
the {\it run I} for inner clusters ($R_{GC}<8 Kpc$; panel $a$) and for outer
clusters ($R_{GC}>8 Kpc$; panel $b$). The line drawn in both panels indicates
the relationship between $C$ and $\log M$ assumed for the initial conditions.
The correlation appears effectively weakened in the \inn sample and well
preserved in the \out one.
\end
\bye

\input basicnew

\newdimen\digitwidth
\setbox0=\hbox{\rm0}
\digitwidth=\wd0

\vsize=6truecm
\hsize=20truecm
\nopagenumbers
\voffset=-6truemm
\tabskip=2em plus1em minus1em
\noindent
{\bf Table 1.} Relevant characteristics of the considered cluster groups
(see Sect. 2).
The subscript "{\it All}" refers to average values calculated including in the
sample both KM (King Model) and PCC (Post Core Collapse) clusters.
\smallskip\noindent
\smallskip\noindent
\halign to
\hsize{
\hfil#\hfil&
\hfil#\hfil&
\hfil#\hfil&
\hfil#\hfil&
\hfil#\hfil&
\hfil#\hfil&
\hfil#\hfil&
\hfil#\hfil\cr
 $Name$ &
$ N_{KM}$ &
$ N_{PCC} $&
$<[Fe/H]>_{All}$ &
$<|Mv|>_{All}$ &
$<Log{\rho_0}>_{KM}$ &
$<Log {r_c}>_{KM}$ &
$<|Z_{GP}|/R_{GC}>_{All}$\cr
DISK~~~&19&~3&$-0.52\pm 0.14$&$7.5\pm 1.2$&$3.96\pm 1.01$&$-0.14\pm 0.32$&$0.28
\pm 0.22$ \cr
INNER~~&34&16&$-1.50\pm 0.32$&$7.6\pm 0.9$&$3.44\pm 0.94$&~$0.03\pm 0.40$&$0.49
\pm 0.27$ \cr
OUTER~~&33&~2&$-1.71\pm 0.33$&$7.4\pm 1.3$&$2.93\pm 1.43$&~$0.20\pm 0.43$&$0.56
\pm 0.26$ \cr
EXTREME&~6&~0&$-1.61\pm 0.31$&$5.9\pm 1.8$&$0.04\pm 0.96$&~$1.00\pm 0.27$&$0.67
\pm 0.18$ \cr
}

\par\noindent
\par\noindent
\par\noindent
\medskip
\vfill\eject
\bye

\input basicnew

\newdimen\digitwidth
\setbox0=\hbox{\rm0}
\digitwidth=\wd0

\vsize=8truecm
\hsize=18truecm
\nopagenumbers
\voffset=-6truemm
\tabskip=2em plus1em minus1em
\noindent
{\bf Table 2.} Spearman's rank correlation coefficients ($s$) for $R_{GC}$ and
$|M_V|$ {\it versus} core parameters. $x_s$ is the probability that the two
considered vectors are actually uncorrelated, so $1-x_s$ can be read
as the confidence
level of the correlation. $x_s$ values as low as $5 \%$ (or lower) are
indicative of {\it significant} correlations.
An asterisk indicates a probability
greater than $20 \%$, that is, the two quantities are {\it not} correlated
(see Sec. 3).
\bigskip\noindent
\halign to
\hsize{
\hfil#\hfil&
\hfil#\hfil&
\hfil#\hfil&
\hfil#\hfil&
\hfil#\hfil&
\hfil#\hfil&
\hfil#\hfil\cr
$~~Name~~$&
$R_{GC}~vs.~Log {r_c}$ &
$R_{GC}~vs.~C$ &
$R_{GC}~vs.~Log{\rho_0}$ &
$|M_V|~vs.~Log {r_c}$ &
$|M_V|~vs.~C$ &
$|M_V|~vs.~Log{\rho_0}$ \cr
}
\halign to
\hsize{
\hfil#\hfil&
\hfil#\hfil&
\hfil#\hfil&
\hfil#\hfil&
\hfil#\hfil&
\hfil#\hfil&
\hfil#\hfil&
\hfil#\hfil&
\hfil#\hfil&
\hfil#\hfil&
\hfil#\hfil&
\hfil#\hfil&
\hfil#\hfil \cr
$~~~~~$ &
$s$ &
$x_s\%$ &
$s$ &
$x_s\%$ &
$s$ &
$x_s\%$ &
$s$ &
$x_s\%$ &
$s$ &
$x_s\%$ &
$s$ &
$x_s\%$ \cr
INNER&~0.41&$<2.5$&-0.22&$<15$~&-0.36&$<5$~~&~0.05&~~*~&~0.14&~~*~&~0.27&~~*~\cr
OUTER&-0.03&~~*~&~0.04&~~*~&~0.01&~~*~&-0.53&$<0.1$&~0.73&$<0.1$&~0.74&$<0.1$\cr
DISK~&~0.46&$<10$~&-0.33&$<15$~&-0.41&$<10$~&-0.42&$<10$~&~0.53&$<5$~~&~0.63&$<0.1$\cr
}

\par\noindent
\par\noindent
\par\noindent
\medskip
\vfill\eject
\bye